\documentclass[12pt]{article}
\title{On Soliton Automorphisms in Massive and Conformal Theories}
\author{Michael M\"uger\thanks{Supported by EU TMR Network `Noncommutative Geometry'} \\ Dipartimento di Matematica, Universit\`{a} di Roma ``Tor Vergata''\\ Via della Ricerca Scientifica, 00133 Roma, Italy\\ Email: mueger@axp.mat.uniroma2.it}

\newlength{\dinwidth}
\newlength{\dinmargin}
\usepackage{amssymb}

\def\1#1{{\bf #1}}
\def\2#1{{\cal #1}}
\def\3#1{{\sl #1}}
\def\4#1{{\tt #1}}
\def\5#1{{\sf #1}}
\def\6#1{{\mathfrak #1}}
\def\7#1{{\mathbb #1}}

\newcommand{\be}{\begin{equation}}
\newcommand{\ee}{\end{equation}}
\newcommand{\ba}{\begin{array}}
\newcommand{\ea}{\end{array}}
\newcommand{\bea}{\begin{eqnarray}}
\newcommand{\eea}{\end{eqnarray}}
\newcommand{\bean}{\begin{eqnarray*}}
\newcommand{\eean}{\end{eqnarray*}}
\newcommand{\nn}{\nonumber}

\newcommand{\impl}{\Rightarrow}
\newcommand{\restr}{\upharpoonright}
\newcommand{\ol}{\overline}

\newcommand{\qed}{\hfill $\blacksquare$}
\newcommand{\qft}{quantum field theory}
\newcommand{\qfts}{quantum field theories}

\newcommand{\npb}{Nucl. Phys. \1B}
\newcommand{\cmp}{Commun. Math. Phys. }

\newcommand{\rmp}{Rev. Math. Phys. }

\newtheorem{defin}{Definition}[section]
\newtheorem{lemma}[defin]{Lemma}
\newtheorem{prop}[defin]{Proposition}
\newtheorem{theorem}[defin]{Theorem}
\newtheorem{coro}[defin]{Corollary}
\newtheorem{conj}[defin]{Conjecture}

\newcommand{\bdefin}{\begin{defin}}
\newcommand{\blemma}{\begin{lemma}}
\newcommand{\bprop}{\begin{prop}}
\newcommand{\btheor}{\begin{theorem}}
\newcommand{\bcoro}{\begin{coro}}
\newcommand{\bconj}{\begin{conj}}
\newcommand{\edefin}{\end{defin}}
\newcommand{\elemma}{\end{lemma}}
\newcommand{\eprop}{\end{prop}}
\newcommand{\etheor}{\end{theorem}}
\newcommand{\ecoro}{\end{coro}}
\newcommand{\econj}{\end{conj}}

\newcommand{\sa}{soliton automorphism}
\newcommand{\poinc}{Poincar\'{e}}
\newcommand{\prf}{{\it Proof. }}
\newcommand{\rem}{{\it Remark. }}
\newcommand{\rems}{{\it Remarks. }}

\newcommand{\sectreset}[1]{\section{#1}\setcounter{equation}{0}}

\begin{document}
\maketitle\noindent

\abstract{For massive and conformal \qfts\ in $1+1$ dimensions with a global gauge group
we consider \sa s, viz.\ automorphisms of the quasilocal algebra 
which act like two different global symmetry transformations on the left and right 
spacelike complements of a bounded region. We give a unified treatment by providing a 
necessary and sufficient condition for the existence and \poinc\ covariance of
\sa s which
is applicable to a large class of theories. In particular, our construction applies to
the QFT models with the local Fock property -- in which case the latter property is the
only input from constructive QFT we need -- and to holomorphic conformal field theories.
In conformal QFT soliton representations appear as twisted sectors, and in a subsequent
paper our results will be used to give a rigorous analysis of the superselection 
structure of orbifolds of holomorphic theories.}

\sectreset{Introduction}
Solitons in massive \qfts\ in $1+1$ dimensions continue to attract the interest of 
quantum field theorists.
Primarily this is due to the fact that they constitute topological excitations of a QFT 
which are non-trivial yet amenable to thorough understanding. They occupy a prominent 
position in the analysis of exactly  soluble classical and quantum models, and there
are strong indications \cite{mue3} that soliton sectors are the only interesting sectors
of massive QFTs in $1+1$ dimensions. The first rigorous approach to the study of 
solitonic sectors in the framework of general QFT was given by Roberts \cite{rob1}, 
and a very general analysis of soliton sectors and their composition structure
has been provided by Fredenhagen \cite{fre1,fre2}. 

As is well known, massive QFTs with a 
spontaneously broken symmetry group give rise to inequivalent vacua and thereby to 
soliton representations. (Of course, spontaneous symmetry breakdown, which occurs only 
for discrete groups \cite{cole}, is not the only possible origin for the existence of 
inequivalent vacua.) Rigorous constructions of soliton sectors and soliton automorphisms
for several models have been given by Fr\"ohlich \cite{fro1,fro2} relying on methods from
algebraic and from constructive \qft.

One aim of this work is to exhibit and exploit the similarities of soliton 
representations of massive and conformal \qfts\ in $1+1$ dimensions with a global
gauge group. Since spontaneous breakdown of inner symmetries
is impossible in conformal theories due to the uniqueness of the vacuum \cite{rob4}
and since left and right spacelike infinity coincide, the role of solitons in CQFT is
necessarily different. Picking a Minkowski space within the conformal covering space
\cite{lm} and restricting the theory to this Minkowski space the soliton condition 
\be\ba{ccr} \pi\restr\2F(W^\2O_{LL}) & = & \pi_0\circ\alpha_g\restr\2F(W^\2O_{LL}),\\
   \pi\restr\2F(W^\2O_{RR}) & = & \pi_0\restr\2F(W^\2O_{RR}), \ea\label{c-sol}\ee
where $W^\2O_{LL}, W^\2O_{RR}$ are the left and right spacelike complements of the
double cone $\2O$, makes sense. It can be shown \cite{mue6} that for $g\ne e$ such a 
representation can not
be unitarily equivalent to $\pi_0$ since it is not locally normal at infinity. Since 
positive energy representations of CQFTs are normal on every double cone \cite{bmt}, 
thus also at infinity, soliton representations do not constitute proper
superselection sectors of the theory $\2F$. In restriction to the fixpoint theory
$\2A=\2F^G$, however, the discontinuity at infinity disappears, $\pi\restr\2A$
being localized in $\2O$:
\be \pi\restr\2A(\2O') = \pi_0\restr\2A(\2O'). \label{DHR}\ee
For this reason the solitons, better known as twisted representations, of the field 
theory $\2F$ are relevant for the superselection structure of the fixpoint theory $\2A$,
cf.\ \cite{dvvv}. The results of this work will be used in a subsequent paper \cite{mue6}
for giving a rigorous analysis of such `orbifold models' (of holomorphic models) and 
clarifying the role of the Dijkgraaf-Witten 3-cocycle $\omega$ and of the twisted
quantum double $D^\omega(G)$. Since the (chiral) Ising model 
\cite{ms,boc1} obviously is not covered by the analysis in \cite{dvvv} although it is an
$\7Z_2$ orbifold model, there must be an implicit assumption in the latter analysis.
Our reconsideration of orbifold models was partially motivated by the desire to clarify 
which properties the triple $(\2F,G,\alpha)$ must possess in order to lead to the 
results of \cite{dvvv}. As it turns out this is just the existence of `twisted sectors' 
in the guise of \sa s, not just of soliton endomorphisms \cite{boc1}.

We briefly recall the framework of local quantum physics \cite{haag,K}. 
We consider a QFT to be given in terms of a net of algebras, i.e.\ a map 
$\2K\ni\2O\mapsto\2F(\2O)$, where $\2K$ is the set of all double cones in Minkowski
space and $\2F(\2O)$ is a $C^*$-algebra. This map being  inclusion preserving
$\2O_1\subset\2O_2 \ \impl \ \2F(\2O_1)\subset\2F(\2O_2)$ we can define the quasilocal
algebra as the inductive limit: $\2F = \ol{\bigcup_{\2O\in\2K}\2F(\2O)}^{\|\cdot\|}$. 
There are commuting automorphic actions on $\2F$ of the \poinc\ group $\2P$ and of a 
locally compact symmetry group $G$ such that 
$\alpha_{\Lambda,x}(\2A(\2O))=\2A(\Lambda\2O+x),\ \forall (\Lambda,x)\in\2P$ and
$\alpha_g(\2F(\2O))=\2F(\2O),\ \forall g\in G$.
When considering observables we require locality, i.e.\ $[\2A(\2O_1),\2A(\2O_2)]=\{0\}$
whenever $\2O_1, \2O_2$ are spacelike to each other. In the presence of fermions we
assume the the usual Bose-Fermi commutation relations w.r.t.\ a $\7Z_2$ grading given by
the automorphism $\alpha_-=\alpha_v$, where $v$ is an element of order two in the center
of $G$.

The quasilocal algebra being automatically simple, all representations are faithful. All
representations $\pi$ we consider are required to be \poinc\ covariant, i.e.\ there is a
unitary representation of $\2P$ with positive energy such that 
$\pi\circ\alpha_x(A)=U_\pi(x)\pi(A)U_\pi(x)^*$. We assume the existence of a  vacuum 
state $\omega_0$ such that the algebras $\pi_0(\2F(\2O))\subset\2B(\2H_0)$ in the GNS 
representation $\pi_0$ on the Hilbert space $\2H$ are weakly closed. Thus $\2F(\2O)$ is 
a $W^*$-algebra and $\alpha_g$ is locally normal. The group of unbroken 
symmetries (w.r.t. $\omega_0$) is defined as
\be G_0=\{g\in G\ |\ \omega_0\circ\alpha_g=\omega_0\}. \ee
If we assume some form of split property (see below) then $G_0$ is automatically compact
when topologized with the strong topology in the representation $\pi_0$ \cite{dl}, and
\cite[Thm.\ 3.6]{dr2} implies $\2F\cap{\2F^{G_0}}'=\7C\11$.
Vacuum representations of local nets are usually required to satisfy Haag duality
\be \pi_0(\2A(\2O)) = \pi_0(\2A(\2O'))' \quad \forall\2O\in\2K,\label{duality}\ee
which for fermionic nets is replaced by twisted duality 
$\pi_0(\2F(\2O))^t = \pi_0(\2F(\2O'))'$ where $X^t=ZXZ^*$ with $Z=\frac{1+iV}{1+i}$,
$V$ being the unitary implementer for $\alpha_-=\alpha_v$ ($v\in G_0$ is automatic 
\cite{rob}).
For many purposes it is sufficient to replace (twisted) Haag duality by wedge duality
$\2R_0(W)^t=\2R_0(W')'\ \forall W\in\2W$, where $\2R_0(W)=\pi_0(\2F(W))''$ and $\2W$
is the set of all wedges, i.e.\ translates of $W_R=\{x\in\7R^2\ | \ x^1\ge |x^0|\}$ and 
the spacelike complement $W_L=W_R'$.

We will be interested in {\it \sa s} of $\2F$, viz. automorphisms 
$\rho^\2O_{g,h}$ which coincide with $\alpha_g$ on the left spacelike complement of a 
double cone $\2O$ (i.e. $W^\2O_{LL}$) and with $\alpha_h$ on the right complement 
($W^\2O_{RR}$). For $g,h\in G_0$ and assuming the existence of a vacuum
representation $\pi_0$ satisfying Haag duality and the split property for wedges (SPW)
such automorphisms can easily be constructed using disorder operators, cf.\ \cite{mue1} 
and the next section. Besides the defining properties of \sa s the 
$\rho^\2O_{g,h}$ obtained in this way have the following remarkable properties:
\begin{itemize}
\item[(a)] The map $G\times G\ni(g,h)\mapsto\rho^\2O_{g,h}\in\mbox{Aut}\,\2F$ is a group 
homomorphism.
\item[(b)] $\rho^\2O_{g,g}=\alpha_g\quad\forall g.$
\item[(c)] $\alpha_k\circ\rho^\2O_{g,h}=\rho^\2O_{kgk^{-1},khk^{-1}}\circ\alpha_k\quad\forall g,h,k$. 
(This property follows from the first two.) In particular, if $G$ is abelian then the 
\sa s commute with the global symmetry. 
\end{itemize}
It is easy to see that families of two-sided \sa s can be equivalently
characterized using only left-handed \sa s. Write
$\rho^\2O_g:=\rho^\2O_{g,e}$. Then properties (a-c) imply:
\begin{itemize}
\item[(A)] $\alpha_k\circ\rho^\2O_g=\rho^\2O_{kgk^{-1}}\circ\alpha_k\quad \forall g,h$. 
\item[(B)] The map $G\ni g\mapsto\rho^\2O_g\in\mbox{Aut}\,\2F$ is a group homomorphism.
\end{itemize}
Conversely, defining 
$\rho^\2O_{g,h}\equiv\alpha_h\circ\rho^\2O_{h^{-1}g}=\rho^\2O_{gh^{-1}}\circ\alpha_h,\  g,h\in G$
one verifies that properties (a-c) follow from (A) and (B).

Postulating the existence of \sa s with the above properties Rehren \cite{khr2} has 
recently derived the modular theoretic assumptions of \cite{n}, where a general proof 
of the cyclic form factor equation is announced. Since these results are quite 
interesting it seems important to understand better when \sa s exist. Our aim will 
thus be to find conditions for the existence of \sa s $\rho^\2O_g$ without appealing to 
the SPW or to $g,h\in G_0$, preferably in such a way that the above properties (A), (B) 
are valid. The paper is organized as follows. Before turning to the general analysis, 
we give two results on solitons in massive theories (characterized by the SPW). In 
particular, in Subsect.\ 2.2 we reconsider the dual net obtained as in \cite{mue1} from 
a massive theory with unbroken abelian symmetry and show that it possesses \sa s 
with all desired properties, which allows to reconstruct the original net.
Sect.\ 3 is the core of the paper and contains the proof of our criterion for the 
existence of \sa s and the proof of their \poinc\ covariance. While 
the existence part is not entirely new, our proof of \poinc\ covariance is and relies
on the uniqueness of soliton sectors up to unitary equivalence, proved in \cite{mue3}.
In Sect.\ 4 we show that the results of Sect.\ 3 apply to all QFT models which possess 
the local Fock property, e.g.\ the $P(\phi)_2$ and $Y_2$ theories.

\sectreset{Solitons in Massive Theories}
\subsection{Soliton Automorphisms from Soliton Sectors}\label{m-sol2}
Given two vacuum representations $\pi_0^L,\pi_0^R$, a representation $\pi$ is said to be
a soliton representation of type $(\pi_0^L,\pi_0^R)$ if it is translation covariant and 
\be \pi\restr\2F(W_{L/R})\cong \pi_0^{L/R}\restr\2F(W_{L/R}) ,\label{solit}\ee
where $W_L,\, W_R$ are arbitrary left and right handed wedges, respectively.
Clearly, a $(\pi_0^L,\pi_0^R)$-soliton representation is locally normal w.r.t. $\pi_0^L$
and $\pi_0^R$. In \cite{schl2} it was shown that for every pair of mutually locally
normal vacuum representations $\pi_0^L,\pi_0^R$ there is a soliton representation of 
type $(\pi_0^L,\pi_0^R)$ if the vacua satisfy Haag duality and the split property for 
wedges (SPW).
Recall that a graded local net $\2F$ with twisted duality satisfies the SPW if for every
double cone $\2O$ there is the following isomorphism of von Neumann algebras:
\be \2R(W^\2O_{LL})\vee\2R(W^\2O_{RR})^t\simeq\2R(W^\2O_{LL})\otimes\2R(W^\2O_{RR})^t. 
\label{SPW}\ee
(This isomorphism is automatically spatial, i.e.\ unitarily implemented.) For free 
massive scalar and Dirac fields the SPW is satisfied, and it will be assumed in this
subsection and the next.

Considering now the case of a broken symmetry let $\pi_g$ be the GNS representation 
corresponding to $\omega_g=\omega_0\circ\alpha_g$. Due to $\pi_g\cong\pi_0\circ\alpha_g$
all ${\pi_g}$'s satisfy Haag duality and the SPW if $\pi_0$ does.
Assuming the existence of an irreducible soliton representation of type $(\pi_g,\pi_0)$
we show that this implies the existence of \sa s of the (abstract) $C^*$-algebra $\2F$,
restricting ourselves to the case of a local net.

\bprop Let $\2F$ be a local net and let $\pi$ be an irreducible soliton representation 
of type $(\pi_g,\pi_0)$,
where one of the (thus both) vacuum representations satisfies Haag duality and the SPW.
Then for each double cone $\2O$ there is an automorphism $\rho^\2O_g$ of $\2F$ such that
$\rho^\2O_g\restr\2F(W^\2O_{LL})=\alpha_g$ and $\rho^\2O_g\restr\2F(W^\2O_{RR})=id$. 
\eprop
\prf By \cite[Thm.\ 4.3]{mue3} $\pi$ automatically satisfies Haag duality, and the SPW 
carries over to $\pi$ by \cite[Thm.\ 5.1]{mue3}. Thus also wedge duality holds 
\cite[Prop.\ 2.5]{mue3}.
Let $\2O\in\2K$. By the soliton criterion there is a representation $\rho_1$ 
on $\2H_0$ equivalent to $\pi$ such that $\rho_1\restr\2F(W^\2O_R)=\pi_0$, where
$W^\2O_R=(W^\2O_{LL})'$. Then we have 
\be \2R(W^\2O_R)=\rho_1(\2F(W^\2O_R))''=\pi_0(\2F(W^\2O_R))''=\2R_0(W^\2O_R),
\label{xyz1}\ee
and wedge duality implies $\2R(W^\2O_{LL})=\2R_0(W^\2O_{LL})$. 
Now, by the soliton criterion $\rho_1$ is equivalent to $\pi_0\circ\alpha_g\cong\pi_g$ 
on every left wedge. Thus there is a unitary $U$ on $\2H_0$ such that 
\be Ad\,U\circ\rho_1\restr\2F(W^\2O_{LL})=\pi_0\circ\alpha_g. \ee
This formula shows that $Ad\,U$ maps the ultraweakly dense subalgebra $\2F(W^\2O_{LL})$
of $\2R(W^\2O_{LL})$ onto another such algebra, and by ultraweak continuity
$U$ acts as an automorphism on $\2R(W^\2O_{LL})$. The SPW for $\pi$ gives rise
to a spatial isomorphism between $\2R(W^\2O_{LL})\vee\2R(W^\2O_{RR})$ and
$\2R(W^\2O_{LL})\otimes\2R(W^\2O_{RR})$ implemented by a unitary $Y^\2O$.
As in \cite{mue1} we define $\tilde{U}=Y^{\2O*}(U\otimes\11)Y^\2O$ which by wedge
duality is contained in $\2R(W^\2O_L)$ and has the same adjoint action on 
$\2R(W^\2O_{LL})$ as $U$. Define
\be \rho_2=Ad\,\tilde{U}\circ\rho_1. \ee
By the localization of $\tilde{U}$ we have
\be \rho_2\restr\2F(W^\2O_{RR})=\rho_1\restr\2F(W^\2O_{RR})=\pi_0, \ee
whereas by construction we have
\be \rho_2\restr\2F(W^\2O_{LL})=\pi_0\circ\alpha_g. \ee
Since $\pi$ satisfies Haag duality we easily see that 
$\pi(\2F(\hat{\2O}))=\pi_0(\2F(\hat{\2O}))$ whenever $\hat{\2O}\supset\2O$. (This was
already observed in \cite[p. 403]{fro2}.) Now the \sa\ is obtained by
$\rho^\2O_g=\pi_0^{-1}\circ\rho_2$. \qed\\
\rem Since the proof involves the SPW it is evident that the interpolation region 
$\2O$ between $\pi_0$ and $\pi_0\circ\alpha_g$ cannot be eliminated.

In \cite{fro1} a soliton {\it representation} for the $\7Z_2$ symmetric $(\phi^4)_2$
theory in the broken phase was constructed by a doubling trick, and the procedure given 
in \cite{schl2} is an abstract version of the former. In both references, however, the 
irreducibility of
the constructed soliton representation is left open, such that the above theorem cannot
be used. Therefore an alternative approach to the construction of \sa s will be developed
in the next section. But before doing so we will give an instructive direct proof of 
the existence of \sa s satisfying conditions (A) and (B) for an interesting special 
class of models considered first in \cite{mue1}.

\subsection{Soliton Automorphisms for Dual Theories}
We start by recalling some results of \cite{mue1}. Again we assume $\2F$ to satisfy 
twisted duality and the SPW. Let $G$ be a group of unbroken, i.e.\ unitarily implemented
symmetries. By the split property $G$ must be strongly compact and second countable 
\cite{dl}, and the Hilbert space $\2H$ is separable. For $g,h\in G$ we define disorder 
operators by
\be\ba{ccc} U^\2O_L(g) = Y^{\2O*} & (U(g)  \otimes  \11) & Y^\2O, \\
   U^\2O_R(h) = Y^{\2O*} & (\11  \otimes  U(h)) &  Y^\2O, \ea\label{disor}\ee
where $Y^\2O$ implements the spatial isomorphism (\ref{SPW}).
One easily verifies $Ad\,U^\2O_L(g)\restr\2F(W^\2O_{LL})=\alpha_g$ and 
$Ad\,U^\2O_L(g)\restr\2F(W^\2O_{RR})=id$
and similarly for $Ad\,U^\2O_R(g)$. Since, as already observed in \cite{fro2},  
$Ad\,U^\2O_L(g)$ maps $\pi_0(\2F(\hat{\2O})), \hat{\2O}\supset\2O$ into itself we 
can obtain \sa s by
\be \rho^\2O_{g,h}=\pi_0^{-1}\circ Ad\,U^\2O_L(g)U^\2O_R(h)\circ\pi_0. \label{sol1}\ee
Using the definition (\ref{disor}) and $Y^\2O U(g)=(U(g)\otimes U(g))Y^\2O$ one can 
verify that the automorphisms (\ref{sol1}) satisfy the properties (a-c).
This construction clearly relies on the existence of the global implementers $U(g)$ of 
$\alpha_g$, which is due to invariance of $\omega_0$. Since these \sa s are unitarily 
implemented in the vacuum representation, they may seem uninteresting from the point of 
view of superselection theory. But, as discussed in \cite{mue1,mue3} they are quite 
useful for elucidating the structure of the fixpoint net $\2A=\2F^G$, which violates 
Haag duality, and its dual net. To this purpose one introduces a nonlocal extension of
the net $\2F$:
\be \hat{\2F}(\2O)=\2F(\2O)\vee U^\2O_L(G)''.\ee
\blemma $\hat{\2F}(\2O)$ is isomorphic to the crossed product 
$\2F(\2O)\rtimes_{\alpha_L} G$, where $\alpha_L(g)=Ad\,U^\2O_L(g)$. \elemma
\prf Recall the result \cite[Appendix]{rob2} according to which 
the action of $G$ on each $\2F(\2O)$ has full spectrum: 
$\Gamma(\alpha\restr\2F(\2O))=\hat{G}$. The same a fortiori holding for the wedge 
algebra $\2R(W^\2O_R)$ and the latter being factorial \cite{dri1},
$\2R(W^\2O_R)\rtimes_\alpha G$ is a factor by \cite[Cor. 6]{glr}. But then
\cite[Corollary 2.3]{haga} gives us 
$\2R(W^\2O_R)\vee U(G)''\simeq\2R(W^\2O_R)\rtimes_\alpha G$. 
Since $\2F(\2O)$ is unitarily equivalent to $\2R(W^\2O_R)\otimes\2R(W^\2O_L)$ and
$U^\2O_L(g)$ to $U(g)\otimes\11$ we are done. \qed\\
\rem In \cite{mue1} this result was obtained only for finite groups, but see also 
\cite{fabio}.

Restricting now to the case of abelian groups $G$, Takesaki duality gives us continuous
actions $\hat{\alpha}^\2O$ of the dual group $\hat{G}$ on all algebras $\hat{\2F}(\2O)$. 
These actions being compatible they give rise to an action of $\hat{G}$ on the quasilocal
algebra $\hat{\2F}$ which is spontaneously broken: 
$\omega_0\circ\hat{\alpha}_\chi\ne\omega_0\ \forall\chi\ne\hat{e}$.
The dual symmetry $\hat{\alpha}$ commuting with the action $\alpha$ of $G$ it acts
on the fixpoint net $\hat{\2A}(\2O)=\hat{\2F}(\2O)^G$, the restriction of which to the
vacuum sector $\2H_0$ was shown to be just the dual net of the fixpoint net 
$\2A\restr\2H_0$. The net $\hat{\2A}\restr\2H_0=(\2A\restr\2H_0)^d$ satisfies Haag 
duality and the SPW and we are in the scenario introduced above. The discrete abelian 
group $\hat{G}$ is countable.

\bdefin A local net $\2B$ satisfying Haag duality and the SPW with a completely broken 
countable abelian symmetry group $K$ is called dual if it arises from a net with unbroken
compact abelian symmetry by the above construction. \edefin
\rem This notation is consistent since $\2B$ is the dual net in the conventional sense of
$\2B=(\2F^G\restr\2H_0)^d$.

In \cite{mue3} it was shown that the representation of $\hat{\2A}$ on the charged
sectors $\2H_\chi\subset\2H$ is the representation of the (unique up to unitary 
equivalence) soliton interpolating between the vacua $\omega_0$ and 
$\omega_\chi\equiv\omega_0\circ\hat{\alpha}_\chi^{-1}$. But we can do better:
\btheor Dual nets admit \sa s satisfying the properties (A), (B). \label{dual}\etheor
\prf Let $\2F$ be the field net with unbroken symmetry $G$ from which $\2B$ arises.
By \cite[Appendix]{rob2} the action of $G$ on each $\2F(\2O)$ has full spectrum, i.e. 
$\forall\2O\in\2K\,\forall\chi\in\hat{G}\,\exists\psi_\chi\in\2U(\2F(\2O)): \alpha_g(\psi_\chi)=\chi(g)\psi_\chi\forall g\in G$.
Due to the split property $\2H$ is separable and $G$ is second countable (called 
separable by many authors). Since the fixpoint algebra $\2A(\2O)$ is properly infinite 
due to the Borchers property we can apply \cite[Prop.\ 20.12]{stra} due to 
Connes and Takesaki which tells us that there is a strongly continuous 
homomorphism $s: K=\hat{G}\rightarrow\2U(\2F(\2O))$ such that 
$\alpha_g(s(\chi))=\chi(g)s(\chi)\ \forall g\in G,\,\chi\in\hat{G}$. (Since $\hat{G}$ 
is discrete in our case continuity is trivial, but the homomorphism property is not.)
Due to the defining properties of disorder operators $\rho_\chi=Ad\,s(\chi)$ implements 
an automorphism of $\hat{\2A}_L$ which is the identity on $W^\2O_{LL}$ and 
$\hat{\alpha}^{-1}_\chi$ on $W^\2O_{RR}$, thus a right \sa. Property (B) is 
fulfilled by construction, and (A) is true since $\rho_\chi$ commutes
with $\hat{\alpha}_\xi$. The \sa s are clearly transportable with intertwiners in 
$\2A=\2F^G=\2B^K$. \qed

Given \sa s satisfying properties (A), (B) one can construct, along the lines of 
\cite{dhr2}, a `crossed product theory' acted upon by the quantum double $D(K)$.
For $K$ abelian this was sketched in \cite{khr2}. Applying this construction to a dual
net in the above sense and restricting to net of $K$ fixpoints one re-obtains the 
original theory with unbroken symmetry under $G=\hat{K}$. In fact, it seems likely that 
every Haag dual net $\2B(\2O)$ with completely broken countable abelian symmetry and with
soliton automorphisms satisfying (A) and (B) is a dual net in the above sense under some
additional conditions. In particular, this should be true if $\2B$ satisfies the SPW
(and thus by \cite[Prop.\ 4.1]{mue3} also property 4 of the following section). We
refrain from going into details since this would lead us to far away from the main 
subject of the present investigation.

In this section we have written $\2B, K$ in order to avoid confusion with the original 
net with from which $\2B$ arises as the dual net. From now on we return to $\2F$ and $G$.

\sectreset{General Approach to Soliton Automorphisms}\label{General}
\subsection{Assumptions and Preliminary Results}
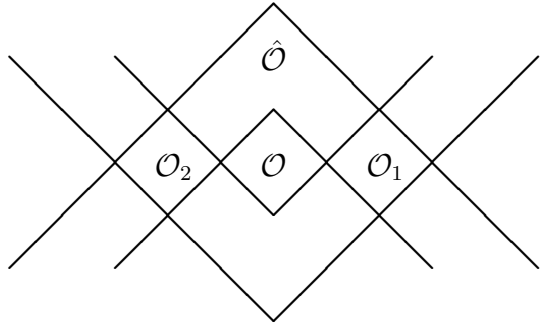
\begin{figure}
\[\ba{c}
\begin{picture}(300,150)(-150,-75)\thicklines
\put(0,60){\line(1,-1){100}}
\put(0,60){\line(-1,-1){100}}
\put(0,-60){\line(1,1){100}}
\put(0,-60){\line(-1,1){100}}
\put(0,20){\line(1,-1){60}}
\put(0,20){\line(-1,-1){60}}
\put(0,-20){\line(1,1){60}}
\put(0,-20){\line(-1,1){60}}

\put(-5,-5){$\2O$}
\put(-5,35){$\hat{\2O}$}
\put(-45,-5){$\2O_2$}
\put(35,-5){$\2O_1$}
\end{picture} 
\ea\]
\caption{Relative spacelike complement of double cones}
\label{rc}\end{figure}

In our considerations of soliton automorphisms we will allow for graded-local nets
since solitons appear in the Yukawa$_2$ model, and since also for the consideration of 
conformal orbifold theories the fermionic case is quite interesting. 
Our assumptions on the field net $\2F$ in the vacuum representation are the following:
\begin{enumerate}
\item (Twisted) Haag duality.
\item Split property (for double cones).
\item The local algebras $\2F(\2O),\ \2O\in\2K$ factors.
\item Minimality of twisted relative commutants, i.e.
\be \2F(\hat{\2O})\wedge{\2F(\2O)^t}'=\2F(\2O_1)\vee\2F(\2O_2) \quad 
  \forall\2O\subset\subset\hat{\2O}, \label{minim}\ee
where $\2O_1, \2O_2$ are related to $\2O, \hat{\2O}$ as in Fig.\ \ref{rc}. 
\item The automorphisms $\alpha_-\restr\2F(\2O)$ are outer for all double cones 
$\2O$.
\end{enumerate}
(In the pure Bose case condition 5 and the twists in condition 1 and 4 disappear.)
We give a few motivating remarks for these
conditions. As to massive theories, all the above properties follow from (twisted) 
duality and the SPW, as was shown for 5 (only for unbroken symmetries) and 3 in 
\cite{mue1} and (in the local case) for 2 and 4 in \cite{mue3}. We refrain from giving 
the easy proofs that the latter properties follow from the SPW also in the fermionic 
case. Free massive scalar and Dirac fields
satisfying (twisted) duality and the SPW, they fulfill our assumptions 1-5. 
Unfortunately, up to now the SPW has not been proven for any interacting theory, but as 
we will see in Sect.\ \ref{locfock}, assumptions 1-5 hold in all models with the local 
Fock property. 

In the case of local conformal fields local factoriality \cite{gafr,bgl} and condition
5 \cite{rob4,mue6} are automatic. The split property is a very weak assumption since it 
follows \cite{gafr} from finiteness of the conformal characters. It is well known that 
condition 1, i.e.\ duality on Minkowski space (or the line) is equivalent to strong 
additivity, which will be shown below to follow from the split property and property 4. 
So we remain with the latter property which is quite restrictive since (in the local 
case) it implies the absence of DHR sectors \cite{dri2,mue3}. Twisted duality on the 
conformal spacetime and local factoriality will be taken for granted also in the 
fermionic case since they can be shown by reconsidering the arguments in \cite{bgl,gafr}.

In \cite{mue3} it was shown that Haag duality and the SPW imply strong additivity. Since
instead of the latter we assume only the conditions 1-5 it is reassuring that we still
have the following.

\blemma The conditions 1-5 imply strong additivity, i.e. 
$\2F(\2O_1)\vee\2F(\2O_2)=\2F(\tilde{\2O})$ whenever the double cones $\2O_1, \2O_2$
are spacelike with a common boundary point, and $\tilde{\2O}$ is the smallest double
cone containing $\2O_1, \2O_2$. \label{sadd}\elemma
\prf Consider $\2O\subset\subset\hat{\2O}$. We will prove 
\be (\2F(\hat{\2O})\wedge{\2F(\2O)^t}')\vee\2F(\2O)=\2F(\hat{\2O}), \label{abc}\ee
which due to condition 4 is equivalent to
$\2F(\2O_1)\vee\2F(\2O)\vee\2F(\2O_2)=\2F(\hat{\2O})$.
The first two algebras being contained in the algebra of the smallest double cone 
containing $\2O_1$ and $\2O$, strong additivity follows. The split property provides 
us with spatial isomorphisms (implemented by the same operator $Y^\Lambda$)
\be\ba{ccccc} \2F(\2O) &\cong& \2F(\2O) &\otimes& \11, \\
   \2F(\hat{\2O})' &\cong& \11 &\otimes& \2F(\hat{\2O})'. \ea\ee
With $\2F(\2O)^t\cong \2F(\2O)_+\otimes\11\,+\,\2F(\2O)_-V\otimes V$ we compute
\be \2F(\hat{\2O})\wedge{\2F(\2O)^t}'\cong{\2F(\2O)^t}'\otimes\2F(\hat{\2O})_+ \, +\,
   \2F(\2O)'\otimes\2F(\hat{\2O})_-. \ee
Condition 3 implies $\2F(\2O)\vee\2F(\2O)'=\2B(\2H)$, and condition 5 entails
$\2F(\2O)\vee{\2F(\2O)^t}'=\2B(\2H)$ as a consequence of
$\2F(\2O)\wedge{\2F(\2O)_+}'=\7C\11$. Thus (\ref{abc}) follows. \qed\\
\rem Note that Haag duality has not been used. To the contrary, in the conformal case the
above result implies Haag duality on Minkowski space when combined with conformal 
duality.

Since the group acts locally normally we have also in the broken symmetry case:
\blemma For every $\Lambda=(\2O,\hat{\2O}),\ \2O\subset\subset\hat{\2O}$ and $g\in G$ 
there is a bosonic unitary implementer $X^\Lambda_g\in\2F(\hat{\2O})$ for 
$\alpha_g\restr\2F(\2O)$. \label{loc-imp0}\elemma
\prf Since the vacuum vector is cyclic and separating for $\2F(\2O)$, the algebra is in 
standard form and there is a unitary representation $X_g$ of $G$ on $\2H$ which 
implements $\alpha\restr\2F(\2O)$, cf.\ \cite[Sect.\ 2]{bdlr} (the construction given 
there coincides with Haagerup's canonical implementation \cite[p.\ 41]{stra}). The
vacuum state being invariant under $\alpha_-$ \cite{rob} there is also a GNS implementer
$V$, with which $X_v$ is easily seen to coincide. Since $v$ is in the center of $G$, the 
$X_g$ commute with $V=X_v$, i.e.\ are bosonic. Using the split property for 
double cones we can define a representation of $G$ in $\2F(\hat{\2O})$ \cite{bdl} by 
$X^\Lambda_g=Y^{\Lambda *}(X_g\otimes\11)Y^\Lambda$ which still implements 
$\alpha\restr\2F(\2O)$. Due to $Y^\Lambda V=(V\otimes V)Y^\Lambda$ also $X^\Lambda_g$ is
bosonic. \qed

In theories with spontaneously broken symmetries the analog of the preceding result for
wedges is false. This fact is responsible for additional difficulties in the treatment
of \sa s as compared to sectors which are localizable in double cones or (left {\it and}
right) wedges.
\blemma Assuming $G$ to be compact, broken symmetries $\alpha_g,\ g\in G-G_0$ act 
non-normally on the algebras of wedges. \label{nonnormal}\elemma
\prf Let $\2A(\2O)=\2F(\2O)^{G}$ and $\2B(\2O)=\2F(\2O)^{G_0}$. As shown in 
\cite[Prop.\ 9]{rob} $\2B(\2O)\subset\2A(W)''$ whenever $\2O\subset W$ and thus 
$\2B(W)''=\2A(W)''$. (In the case $G_0=\{e\}$ this implies that the subnet $\2A$ still 
satisfies wedge duality on the Hilbert space $\2H$.) Assuming that $\alpha_g$ acts 
normally on $\2F(W)$, it acts triviality on $\2B(W)$ since by definition it is trivial 
on the ultraweakly dense subalgebra $\2A(W)$. Then by translation invariance $\alpha_g$ 
acts trivially on $\2B$. Now by \cite[Thm.\ 3.6 b)]{dr2} every automorphism of $\2F$
which acts trivially on $\2B$ is a gauge automorphism $\alpha_g$ with $g\in G_0$. \qed\\
\rems 1. We emphasize that \cite[Thm.\ 3.6]{dr2} is true also in $1+1$ dimensions
with the possible exception of twisted duality for $\2F$, which we do not need to prove
anyway since it is one of our axioms. \\
2. The only step where compactness of $G$ is used is the argument in 
\cite[Prop.\ 9]{rob} leading to $\2B(\2O)\subset\2A(W)''$ for $\2O\subset W$.
The latter is seen without difficulty to work also for locally compact abelian groups
acting integrably on the algebras $\2F(\2O)$. This is due to the fact 
\cite[Cor.\ 21.3]{stra} that $\2F(\2O)$ is generated by the operators transforming 
under the action of $G$ by multiplication with a character. Finally, since $\2H$ is
separable due to the split property, the integrability property follows (for separable 
$G$) by an application of \cite[Prop.\ 20.12]{stra} if we assume that for
every $\gamma\in\hat{G}$ there is a field operator in $\2F(\2O)$ transforming according
to $\gamma$. While for non-compact groups the argument in \cite[Appendix]{rob2} does not
work, this assumption is physically very reasonable. It is satisfied, e.g., in the
sine-Gordon model where $G=\7Z$. Lemma \ref{nonnormal} will not be used in this paper.

The considerations in the sequel are considerably more transparent in the case of purely
bosonic, i.e.\ local nets. For this reason I prefer as in \cite{mue4} to treat the pure 
Bose case first in order to avoid confusion by the inessential complications of the 
general Bose/Fermi situation.

\subsection{Existence of Soliton Automorphisms: Bose Case}
The following is a version of a well-known result from \cite{dopl}. It follows by 
plugging condition 4 into \cite[Prop.\ 4.2]{mue4}, but incorporates simplifications
using strong additivity. We state the proof mainly for reference purposes in the next
subsection were we treat the fermionic case.
\blemma Let $\2O\mapsto\2F(\2O)$ fulfill conditions 1-4.
Then every locally normal endomorphism $\rho$ of the quasilocal algebra $\2F$ satisfying
$\rho\restr\2F(\2O')=id$ for some $\2O\in\2K$ is an inner endomorphism of $\2F$, i.e. a 
direct sum of copies of the identity morphism. \label{do-lemma}\elemma
\prf Let $\rho$ be localized in $\2O$ and choose a double cones $K$ fulfilling
$\2O\subset\subset K\subset\subset\hat{\2O}$. Thanks to the
split property there exist type I factors $M_1, M_2$ such that
\be \2F(\2O)\subset M_1\subset\2F(K)\subset M_2\subset\2F(\hat{\2O}).\ee
By Haag duality $\rho$ maps $\2F(\tilde{\2O})$ into itself whenever 
$\2O\subset\tilde{\2O}$, in particular $\rho(M_1)\subset\2F(K)$. Being localized in 
$\2O$, $\rho$ acts trivially on
$M_1'\cap\2F(\hat{\2O})\subset\2F(\2O)'\cap\2F(\hat{\2O})=\2F(\hat{\2O}\cap\2O')$,
where we have used condition 4. This implies
\be \rho(M_1)\subset (M_1'\cap\2F(\hat{\2O}))'\cap M_2 \subset (M_1'\cap M_2)'\cap M_2
    = M_1 ,\ee
the last identity following from $M_2$ being type I. Thus $\rho$ restricts to an
endomorphism of $M_1$. Now every normal endomorphism of a type I factor is inner 
\cite[Cor. 3.8]{lo1}, i.e. there is a (possibly infinite) family of isometries 
$V_i\in M_1,\ i\in I$ with 
$V_i^* V_j=\delta_{i,j},\sum_{i\in I}V_i V_i^*=\11$ such that 
$\rho\restr M_1=\eta\restr M_1$, where
\be \eta(A)\equiv\sum_{i\in I} V_i \, A \, V_i^*  \label{inner}\ee
is well-defined on $\2B(\2H)$, the sums over $I$ being understood in the strong sense.
Again $\rho$ acts trivially on $\2F(\2O)'\cap M_1\subset\2F(\2O)'\cap \2F(\hat{\2O})$,
which implies $V_i\in(\2F(\2O)'\cap M_1)'\cap M_1=\2F(\2O)\ \forall i\in I$. 
Therefore in addition to $\rho=\eta$ on $\2F(\2O)$ we have $\rho=\eta=id$ on 
$\2F(\2O')$. But now local normality of $\rho$ and strong additivity, cf.\ Lemma 
\ref{sadd}, imply $\rho=\eta$ on all local algebras, thus on $\2F$. \qed

Now we can identify a necessary condition for the existence of \sa s.
\bprop Let $\2F$ satisfy the assumptions 1-4. Let $\2O\subset\subset\hat{\2O}$ and 
let $\2O_1, \2O_2$ be as in Fig.\ \ref{rc}. Let $\rho^{\2O_1}_g, \rho^{\2O_2}_g$ be 
locally normal \sa s. Then there is a unitary $U^\Lambda_g\in\2F(\hat{\2O})$, unique 
up to a phase, such that
$\rho^{\2O_1}_g\circ(\rho^{\2O_2}_g)^{-1}=Ad\,U^\Lambda_g$. Furthermore, 
$Ad\,U^\Lambda_g\restr\2F(\2O)=\alpha_g$ and $Ad\,U^\Lambda_g$ leaves 
$\2F(\2O_1)$ and $\2F(\2O_2)$ stable. 
\label{loc-imp4}\eprop
\prf It is obvious from the definition of \sa s that 
$\gamma=\rho^{\2O_1}_g\circ(\rho^{\2O_2}_g)^{-1}$ acts trivially on 
$\2F(\hat{\2O}')$ and as $\alpha_g$ on $\2F(\2O)$. Now by Lemma \ref{do-lemma}
this implies that $\gamma=Ad\,U^\Lambda_g$ where $U^\Lambda_g\in\2U(\2F(\hat{\2O}))$. 
$U^\Lambda_g$ is unique up to a phase by irreducibility of $\2F$. Now, 
$\rho^{\2O_1}_g$ acts as an inner symmetry on $\2F(\2O_2)$ and leaves $\2F(\2O_1)$ 
stable due to Haag duality. Arguing similarly for $\rho^{\2O_2}_g$ we see that 
$\gamma=Ad\,U^\Lambda_g$ leaves $\2F(\2O_1)$ and $\2F(\2O_2)$ stable. \qed\\
\rems 1. This result is the analog in the broken symmetry case of \cite[Lemma 2.3]{mue1}
which stated the uniqueness of disorder operators up to localized unitaries.\\
2. Given a family of \sa s we see that transportability is automatic.
This fact will play a crucial role in our proof of \poinc\ covariance.

By Lemma \ref{loc-imp0} local implementers exist also in the case of broken 
symmetry. Whereas a local implementer $X^\Lambda_g$ acts as an automorphism on 
$\2F(\hat{\2O})\cap\2F(\2O)'=\2F(\2O_1)\vee\2F(\2O_2)$,
the above property of leaving $\2F(\2O_1)$ and $\2F(\2O_2)$ {\it separately}
stable is stronger. We will now show the converse, viz. the existence of local
implementers with this additional property implies the existence of soliton
automorphisms.

\bprop Let $\2F$ satisfy assumptions 1-4. If for some $g\in G$ and all 
$\Lambda=(\2O,\hat{\2O}),\ \2O\subset\subset\hat{\2O}$ there is 
$U^\Lambda_g\in\2U(\2F(\hat{\2O}))$ such that $Ad\,U^\Lambda_g\restr\2F(\2O)=\alpha_g$
and such that $Ad\,U^\Lambda_g$ restricts to automorphisms of $\2F(\2O_i), i=1,2$
then there are locally normal left \sa s $\rho^\2O_g\ \forall \2O\in\2K$. 
\label{sol2}\eprop
\prf Choose two double cones $\2O_1$ and $\2O_2<<\2O_1$ (i.e.\ 
$\2O_2\subset\subset W^{\2O_1}_{LL}$). Let 
$\hat{\2O}_2$ be the smallest double cone containing $\2O_1$ and $\2O_2$, and let
$\2O=\hat{\2O}_2\cap\2O_1'\cap\2O_2'$. Then there is a unitary $z_2$ in 
$\2F(\hat{\2O}_2)$ implementing $\alpha_g$ on $\2F(\2O)$ and implementing 
automorphisms of $\2F(\2O_1), \2F(\2O_2)$. Our aim will be to construct a \sa\
which acts like $Ad\,z_2$ on $\2F(\2O_1)$ and like $\alpha_g$ on the left complement of 
$\2O_1$. Considering thus another double cone $\2O_3<<\2O_2$ we denote by $\hat{\2O}_3$ 
the smallest double cone containing $\2O_1$ and $\2O_3$ (thus also $\2O_2$) and by 
$\tilde{\2O}_3$ the smallest double cone containing $\2O_2$ and $\2O_3$. Again there is 
$\tilde{z}_3$ implementing $\alpha_g$ on the double cone between $\2O_1$ and $\2O_3$
and acting on $\2F(\2O_1), \2F(\2O_3)$ by automorphisms.
Now, $X=\tilde{z}_3^* z_2\in\2F(\hat{\2O}_3)$ commutes with $\2F(\2O)$, which by 
condition 4 implies $X\in\2F(\2O_1)\vee\2F(\tilde{\2O}_3)$. This algebra being 
isomorphic to the tensor product $\2F(\2O_1)\otimes\2F(\tilde{\2O}_3)$ and the adjoint 
action of $X$ implementing an automorphism of $\2F(\2O_1)$, the lemma below implies 
$X=X_1X_3$ where $X_1\in\2U(\2F(\2O_1)), X_3\in\2U(\2F(\tilde{\2O}_3))$. Defining 
$z_3=\tilde{z}_3 X_1$ we have $Ad\,z_3\restr\2F(\2O_1)=Ad\,z_2\restr\2F(\2O_1)$ and 
$z_3$ still implements $\alpha_g$ on the double cone between $\2O_3$ and $\2O_1$.
Let  now $\2O_n$ be a sequence of double cones tending to left spacelike infinity. More 
precisely, for every $\2O\in\2K$ there is a $N\in\7N$ such that 
$\2O_n<\2O\ \forall n>N$. In the above way we can construct operators $z_n, n\in\7N$ 
implementing $\alpha_g$ on the double cone between $\2O_1$ and $\2O_n$ and such that 
$Ad\,z_n\restr\2F(\2O_1)=Ad\,z_2\restr\2F(\2O_1)\ \forall n$. Defining 
\be \rho^{\2O_1}_g(A)=\|\cdot\|-\lim_{i\rightarrow\infty} z_i A z_i^* ,\label{sol3}\ee
it is clear that $\rho^{\2O_1}_g$ is a locally normal \sa. \qed
\blemma Let $M_1, M_2$ be factors and let $U\in\2U(M_1\otimes M_2)$ be such that
$UM_1\otimes\11 U^*=M_1\otimes\11$. Then $U=U_1\otimes U_2$ where $U_i\in\2U(M_i)$.
\elemma
\prf Due to factoriality we have $\11\otimes M_2=(M_1\otimes M_2)\cap(M_1\otimes\11)'$.
Thus $\alpha=Ad\,U$ stabilizes also $\11\otimes M_2$ and factorizes: 
$\alpha=\alpha_1\otimes\alpha_2$. $\alpha$ being inner, the same holds for
$\alpha_1, \alpha_2$ by \cite[Prop. 17.6]{stra}.
Thus there are unitaries $U_i\in M_i$ such that $\alpha=Ad\,U_1\otimes U_2$. 
Since we are dealing with factors the inner implementer is unique up to a phase and 
$U_1, U_2$ can be chosen such that $U=U_1\otimes U_2$. \qed\\
\rem Obviously the proof of the above proposition is in the spirit of Roberts' local 
cohomology theory. 

Up to now have established a necessary and sufficient criterion for the existence of
arbitrarily localizable \sa s. We conclude this subsection by giving sufficient criteria
in terms of the localized implementers $U^\Lambda_g$ for the \sa s to satisfy conditions
(A), (B).
\bprop If for every $\Lambda=(\2O,\hat{\2O})$ there are localized implementers 
$U^\Lambda_g$ which besides the properties required in Prop.\ \ref{sol2} satisfy 
$\alpha_k(U^\Lambda_g)=U^\Lambda_{kgk^{-1}}$ then there are \sa s satisfying property 
(A). If there are $U^\Lambda_g$'s such that $U^\Lambda_gU^\Lambda_h=U^\Lambda_{gh}$ then
there are \sa s satisfying property (B). If there are $U^\Lambda_g$'s satisfying both 
conditions then there are \sa s fulfilling (A) and (B). \label{AB}\eprop
\prf By the definition of \sa s it is obvious that $\rho^\2O_g\rho^\2O_h=\rho^\2O_{gh}$
and $\alpha_k\circ\rho^\2O_g=\rho^\2O_{kgk^{-1}}\circ\alpha_k$ are satisfied in 
restriction to $\2F(\2O')$. By strong additivity and local normality of the \sa s it 
suffices to prove the above relations for $\2F(\2O)$. But since there are \sa s such
that $\rho^{\2O_1}_g\restr\2F(\2O_1)=Ad\,U^\Lambda_g$ for some $\Lambda$, the claimed 
implications are obvious consequences of the assumptions. \qed\\
\rem The conditions given above are stronger than necessary, since the adjoint action of
$U^\Lambda_g$ on $\2O_2$ does not leave traces in the $\rho^{\2O_1}_g$ which is 
constructed in the theorem. For our purposes the above result is sufficient. Anyhow,
a detailed investigation of when properties (A), (B) can be satisfied, part of which
will be found in \cite{mue6}, would have to be cohomological.

\subsection{Existence of Soliton Automorphisms: General Case}\label{fermi}
In the local case condition 4 immediately gives us the relative commutant of two double 
cone algebras. In the general case we instead have the following 
\blemma With the notation of Fig.\ \ref{rc} we have
\be \2F(\hat{\2O})\wedge\2F(\2O)'= (\2F(\2O_1)\vee\2F(\2O_2))_+ \ + \
   (\2F(\2O_1)\vee\2F(\2O_2))_- \, U^\Lambda_v,
\label{rc5}\ee
where $U^\Lambda_v$ is an arbitrary Bose unitary in $\2F(\hat{\2O})$ implementing
$\alpha_-$ on $\2F(\2O)$. \label{rc-e}\elemma
\prf The Bose part of the left hand side in (\ref{rc5}) is given by 
$\2F(\hat{\2O})_+\wedge\2F(\2O)'=\2F(\hat{\2O})_+\wedge{\2F(\2O)^t}'$, where we have
used $(\2F(\hat{\2O})_+)^t=\2F(\hat{\2O})_+$. Using condition 4 we see that for the Bose
parts (\ref{rc5}) is correct. As to the Fermi part we know by Lemma \ref{loc-imp0} that
a $U^\Lambda_v$ as needed exists, and it is easy to see that 
$(\2F(\2O_1)\vee\2F(\2O_2))_-U^\Lambda_v\subset\2F(\hat{\2O})_-\wedge\2F(\2O)'$.
Conversely, if $X\in\2F(\hat{\2O})_-\wedge\2F(\2O)'$ then 
$XY\in\2F(\hat{\2O})_+\wedge\2F(\2O)'$ for a unitary 
$Y\in(\2F(\2O_1)\vee\2F(\2O_2))_-U^\Lambda_v$. Since $\2F(\hat{\2O})\wedge\2F(\2O)'$
is an algebra it contains also $X$. \qed

In the sequel an automorphism will be called {\it even} if it commutes with $\alpha_-$,
i.e.\ respects the $\7Z_2$ grading.
We begin by reconsidering Prop.\ \ref{sol2}, assuming the assumptions made there plus
condition 5. All geometrical notions are as in the proof of Prop.\ \ref{sol2} and
let $z_2$ be as defined there, but in addition we need to assume that $Ad\,z_2$ acts on
$\2F(\2O_1),\2F(\2O_2)$ by even automorphisms.
By the split property $\2F(\2O_1)\vee\2F(\2O_2)^t$ is isomorphic to the tensor product,
thus a factor. That also $\2F(\2O_1)\vee\2F(\2O_2)$ is factorial is proved by exactly
the same argument as in \cite[Cor.\ 3.13]{mue1}. (Here condition 5 is used.)
By Lemma \ref{loc-imp0} there is a Bose implementer $V_1$ of $\alpha_-\restr\2F(\2O_1)$
which commutes with $\2F(\2O_2)$. Using the split property it is easy to see that
$Ad\,V_1$ acts as an automorphism on $(\2F(\2O_1)\vee\2F(\2O_2))_+$ with fixpoint algebra
$\2F(\2O_1)_+\vee\2F(\2O_2)_+$. Since $Ad\,z_i\restr\2F(\2O_1)$ is an even automorphism,
$V_1XV_1^*\,X^*$ commutes with $\2F(\2O_1)$ and the same holds trivially for 
$\2F(\2O_2)$. Thus by factoriality of $\2F(\2O_1)\vee\2F(\2O_2)$ we have 
$V_1XV_1^*=\pm X$. In case the minus sign occurs we replace $z_2$ by 
$z_2\,Y_1Y_2$ with $Y_i\in\2F(\2O_i), i=1, 2$ Fermi unitaries. Since $Y_1Y_2$ is bosonic
the required implementation properties of $z_2$ are not affected. We may thus assume 
that $z_2$ commutes with $V_1$, and the same holds for the $\tilde{z}_n$'s. Reconsidering
$X=\tilde{z}_3^*z_2\in\2F(\hat{\2O}_3)_+\cap\2F(\2O)'$, the Bose part of Lemma \ref{rc-e}
implies $X\in(\2F(\2O_1)\vee\2F(\tilde{\2O}_3))_+$. By the above we can assume that $X$
commutes with $V_1$ which yields $X\in\2F(\2O_1)_+\vee\2F(\tilde{\2O}_3)_+$. The latter
algebra being isomorphic to a tensor product the rest of the proof works as in the Bose
case and (\ref{sol3}) defines an even \sa\ $\rho^{\2O_1}_g$. Since all the $z_n$'s 
commute with $V_1$, we have $\rho^{\2O_1}_g(V_1)=V_1$. Returning to our standard notation
for localized implementers this means that $\rho^{\2O_1}_g$ is bosonic in the following
sense.
\bdefin Let $\rho$ be an endomorphism of $\2F$ which is localized in $\2O$ in the 
usual or solitonic sense. Let $U_v^\Lambda\in\2F(\hat{\2O})_+$ be a localized bosonic 
implementer for $\alpha_-\restr\2F(\2O)$. Then $\rho$ is called bosonic if 
$\rho(U_v^\Lambda)=U_v^\Lambda$. \edefin
\rems 1. This definition is independent of the choice of $U_v^\Lambda$ since for another 
choice $\tilde{U}_v^\Lambda\in\2F(\hat{\2O})_+$ we have 
$\tilde{U}_v^\Lambda U_v^{\Lambda*}\in\2F(\hat{\2O})_+\wedge\2F(\2O)'=\2F(\hat{\2O}\cap\2O')_+$,
on which $\rho$ acts trivially.\\
2. If $\rho$ is implemented (on a bigger Hilbert space) by a multiplet of field 
operators $\Psi_i$ then $\rho$ is bosonic iff the $\Psi_i$ commute with $U^\Lambda_v$,
i.e.\ are Bose fields.

Let conversely the soliton automorphisms $\rho_g^{\2O_1},\rho_g^{\2O_2}$ considered in 
Prop.\ \ref{loc-imp4} in addition be even and bosonic. Then
$\gamma=\rho_g^{\2O_1}\circ(\rho_g^{\2O_2})^{-1}$ clearly is a bosonic even endomorphim 
localized in the double cone $\hat{\2O}$ of Fig.\ \ref{rc}. As shown in 
\cite{mue4} evenness implies that $\gamma$ still 
maps $\2F(\hat{\2O})$ into itself and in Lemma \ref{do-lemma} we have 
$\gamma(M_1)\subset\2F(K)$. Another change in the latter Lemma is due to the fact that
in the case with fermions the relative 
commutant $\2F(\2O)'\cap\2F(\hat{\2O})$ appearing in the proof is not given by 
$\2F(\hat{\2O}\cap\2O')$, but instead as in Lemma \ref{rc5}. (Note that the $\2O$ in 
Lemma \ref{do-lemma} corresponds to the above $\hat{\2O}$, wheras the 
$\hat{\2O}$ appearing there has nothing to do with that above!) Yet 
$\gamma$ still acts trivially on the relative commutant, since it is bosonic and 
thus acts trivially on the $U^\Lambda_v$ in Lemma \ref{rc5}. Thus the conclusion of 
Lemma \ref{do-lemma} holds and we have a Bose unitary $U^\Lambda_g\in\2F(\hat{\2O})$ 
such that $\gamma=Ad\,U^\Lambda_g$. Clearly $U^\Lambda_g$ implements
$\alpha_g\restr\2F(\2O)$ and even automorphisms of $\2F(\2O_1),\2F(\2O_2)$.

We summarize the results of the preceding discussion in the following
\btheor Let $\2F$ satisfy conditions 1-5. Then the following are equivalent
\begin{itemize}
\item[(i)] There are [bosonic even] \sa s $\rho^\2O_g\ \ \forall g\in G,\2O\in\2K$.
\item[(ii)] For every $g\in G, \Lambda=(\2O,\hat{\2O}), \2O\subset\subset\hat{\2O}$ 
there is a [bosonic] unitary implementer $U^\Lambda_g\in\2F(\hat{\2O})$ for 
$\alpha_g\restr\2F(\2O)$ whose adjoint action implements [even] automorphisms of 
$\2F(\2O_i),i=1,2$.
\end{itemize}
(In the pure Bose case omit the words within square brackets.)
Prop.\ \ref{AB} remains true in the Bose/Fermi case. \label{main0}\etheor

\subsection{\poinc\ Covariance}\label{pcov}
Up to now we have seen that under certain conditions there are even soliton automorphism
which are transportable with Bose intertwiners.
In \cite[Thm.\ 5.2]{gl} it was proved that every transportable sector which is 
localizable in wedges and has finite statistics is \poinc\ covariant provided certain 
conditions on the net are satisfied. Unfortunately this result cannot applied
since soliton automorphisms typically are non-normal on the algebra of the wedge in which
they are localized, cf.\ Lemma \ref{nonnormal}. We will thus adopt another approach, always assuming the conditions 1-5 on the net $\2F$.
\blemma Let $\rho^\2O_g$ be a bosonic even \sa. Then for every $(\Lambda,x)\in\2P$ 
and a double cone $\hat{\2O}$ containing $\2O$ and $\Lambda\2O+x$ the automorphism
$\beta_{(\Lambda,x)}$ defined by
\be \alpha_{(\Lambda,x)}\circ\rho^\2O_g\circ\alpha_{(\Lambda,x)}^{-1}(F) =
  \beta_{(\Lambda,x)} \circ\rho^\2O_g(F)\quad\forall F\in\2F \label{cov1}\ee
is implemented by a unitary $Z(\Lambda,x)\in\2F(\hat{\2O})$ which is determined up
to an arbitrary phase. The map
\be \2P\ni(\Lambda,x)\mapsto\alpha^\rho_{(\Lambda,x)}\equiv\beta^{-1}_{(\Lambda,x)}\circ
   \alpha_{(\Lambda,x)} \ee
is a group homomorphism. \elemma
\prf $\alpha_{(\Lambda,x)}\circ\rho^\2O_g\circ\alpha_{(\Lambda,x)}^{-1}$ is a bosonic 
even \sa\ for $g$ localized in $\Lambda\2O+x$. Thus by Lemma \ref{do-lemma} and the 
discussion in the preceding subsection there is a unitary $Z(\Lambda,x)$ in 
$\2F(\hat{\2O})$ with $\hat{\2O}$ as above, such that (\ref{cov1}) holds with 
$\beta_{(\Lambda,x)}=Ad\,Z(\Lambda,x)$. By irreducibility of the quasilocal algebra 
$Z(\Lambda,x)$ is unique up to a phase. Now 
$\beta^{-1}_{(\Lambda,x)}\circ\alpha_{(\Lambda,x)}=\rho_g^\2O\circ\alpha_{(\Lambda,x)}\circ(\rho^\2O_g)^{-1}$,
and the group property of $\alpha^\rho$ is obvious. \qed

\bprop The automorphisms $\alpha^\rho_{(\Lambda,x)}$ of the preceding lemma
extend uniquely to $\2B(\2H)$. 
The action $\2P\ni(\Lambda,x)\mapsto\alpha^\rho_{(\Lambda,x)}\in\mbox{Aut}\,\2B(\2H)$ is 
continuous. \eprop
\rem The natural topology on the automorphism group of a von Neumann algebra $A$ is
the u-topology, viz.\ the topology defined by norm convergence on the predual, since it 
turns $\mbox{Aut}\,A$ into a topological group. For a type I factor $\2B(\2H)$ this
topology coincides with the p-topology, which is the restriction to 
$\mbox{Aut}\,\2B(\2H)$ of the pointwise weak topology, cf.\ \cite[pp.\ 41-43]{stra}. \\
\prf The extendibility statement is obvious since $\alpha^\rho_{(\Lambda,x)}$ is 
unitarily implemented, and uniqueness follows since $\2F$ is irreducible. It clearly 
suffices to prove continuity for a neighbourhood $\2V\subset\2P$ of the unit element 
$e=(\11,0)$. Let $\2O,\hat{\2O}$ be double cones such that 
$\Lambda\2O+x\subset\hat{\2O}\ \forall (\Lambda,x)\in\2V$. Then by the lemma we have 
\be \beta_{(\Lambda,x)}=\alpha_{(\Lambda,x)}\circ\rho_g^\2O\circ\alpha^{-1}_{(\Lambda,x)}
   \circ(\rho_g^\2O)^{-1} \ee
with $\beta_{(\Lambda,x)}=Ad\,Z(\Lambda,x)$ and $Z(\Lambda,x)\in\2F(\hat{\2O})$.
Thus the $\beta_{(\Lambda,x)}, (\Lambda,x)\in\2V$ are inner automorphisms of every
algebra which contains $\2F(\hat{\2O})$. Furthermore, for every localized $F$ the map
$(\Lambda,x)\mapsto\beta_{(\Lambda,x)}(F)$ is strongly continuous since $\rho_g^\2O$ is
locally normal and $\alpha$ is a continuous action. Let now 
$\tilde{\2O}\supset\supset\hat{\2O}$ be another double cone and let $M$ be a type I
factor such that $\2F(\hat{\2O})\subset M\subset\2F(\tilde{\2O})$, the existence of
which is guaranteed by the split property. By the above
$\beta_{(\Lambda,x)}, (\Lambda,x)\in\2V$ acts continuously on $\2F(\tilde{\2O})$, thus 
on $M$ and trivially on $M'\subset\2F(\hat{\2O})'$. But 
$\2B(\2H)=M\vee M'\simeq M\otimes M'$ implies that
$\beta_{(\Lambda,x)}, (\Lambda,x)\in\2V$ acts continuously on $\2B(\2H)$. (We have used
that, w.r.t.\ the u-topologies, $\alpha_\iota\rightarrow\alpha$ implies 
$\alpha_\iota\otimes\mbox{id}\rightarrow\alpha\otimes\mbox{id}$ and that 
$M, M', \2B(\2H)$ are type I, such that the above remark on the topologies applies.)
Since $Aut\,\2B(\2H)$ is a topological group also the action 
$(\Lambda,x)\mapsto\alpha^\rho_{(\Lambda,x)}=\beta^{-1}_{(\Lambda,x)}\circ\alpha_{(\Lambda,x)}$
of $\2P$ on $\2B(\2H)$ is continuous for $(\Lambda,x)\in\2V$, thus for all of $\2P$. \qed

\btheor The \sa\ $\rho^\2O_g$ is \poinc\ covariant, i.e.\ in the sector 
$\pi_0\circ\rho^\2O_g$ there is a strongly continuous representation 
$U^\rho(\Lambda,x)$ of the \poinc\ group with positive energy such that 
$Ad\,U^\rho(\Lambda,x)\circ\pi_0\circ\rho^\2O_g=\pi_0\circ\rho^\2O_g\circ\alpha_{(\Lambda,x)}$.
\label{p-covar}\etheor
\prf By the preceding results we are in a position to apply a result of Kallman and 
Moore \cite[Thm.\ 15.16]{stra} 
according to which every continuous one parameter group of inner automorphisms of a von 
Neumann algebra $M$ with separable predual is implemented by a strongly continuous 
unitary representation in $M$. (Since $\2B(\2H)$ is a factor the proof by Hansen, 
reproduced in \cite[p.\ 218]{stra}, is sufficient for our purposes.) Thus we have
\be \alpha_\Lambda^\rho=Ad\,e^{i\Lambda K^\rho},\quad
  \alpha^\rho_{+,a}=Ad\,e^{iaP^\rho_+},\quad
  \alpha^\rho_{-,b}=Ad\,e^{ibP^\rho_-}, \label{gener}\ee
where $\alpha^\rho_{\pm}$ are the lightlike translations and $K^\rho, P^\rho_+, P^\rho_-$
are self-adjoint operators on $\2H$. From now on we omit the superscript $\rho$. 
The unitary implementer of an automorphism being unique up to a phase
the commutation relation 
$\alpha_\Lambda\circ\alpha_{+,a}=\alpha_{+,e^\Lambda a}\circ\alpha_\Lambda$ in $\2P$
together with (\ref{gener}) implies
$e^{i\Lambda K}e^{iaP_+}=c(a,\Lambda)\ e^{ie^\Lambda aP_+}e^{i\Lambda K}$, where $c$ 
is a continuous phase-valued function satisfying
$c(a_1+a_2,\Lambda)=c(a_1,\Lambda)c(a_2,\Lambda)$ and
$c(a,\Lambda_1+\Lambda_2)=c(a,\Lambda_1)c(e^{\Lambda_1}a,\Lambda_2)$. Together with 
continuity the first equation implies $c(a,\Lambda)=e^{iad(\Lambda)}$, where $d$ 
satisfies $d(\Lambda_1+\Lambda_2)=d(\Lambda_1)+e^{\Lambda_1}d(\Lambda_2)$. The left hand
side of this equation being symmetric in $\Lambda_1,\Lambda_2$ we have
$d(\Lambda_1)+e^{\Lambda_1}d(\Lambda_2)=d(\Lambda_2)+e^{\Lambda_2}d(\Lambda_1)$, which
for $\Lambda_2\ne 0$ gives
$d(\Lambda_1)=(e^{\Lambda_1}-1)d(\Lambda_2)/(e^{\Lambda_2}-1)=A(e^{\Lambda_1}-1)$. 
Proceding similarly for the other commutation relations we thus have
\bea e^{i\Lambda K}e^{iaP_+} &=& e^{iAa(e^\Lambda-1)}\ e^{ie^\Lambda aP_+}e^{i\Lambda K},
  \label{R1}\\
   e^{i\Lambda K}e^{ibP_-} &=& e^{iBb(e^{-\Lambda}-1)}\ e^{ie^{-\Lambda}bP_-}
     e^{i\Lambda K},\\
   e^{iaP_+}e^{ibP_-} &=& e^{iCab}\ e^{ibP_-}e^{iaP_+}, \label{R3}\eea
where $A, B, C\in\7R$ are independent. By differention we obtain the commutation 
relations
\be i[K,P_+]=P_++A\11,\ \ i[K,P_-]=-(P_-+B\11),\ \ i[P_+,P_-]=C\11, \ee
which we need not make precise.
Now, the generators $K, P_+, P_-$ are determined by (\ref{gener})
only up to addition of a multiple of $\11$. Using this freedom we can replace $P_+$ by
$P_++A\11$, which removes the factor $e^{iAa(e^\Lambda-1)}$ from (\ref{R1}). This fixes 
$P_+$ uniquely and achieves that the Lorentz group acts on $P_+$ as dilatations:
$e^{i\Lambda K}P_+e^{-i\Lambda K}=e^\Lambda P_+$. Thus the spectrum of $P_+$ is one of
the sets $\{0\}, [0,\infty), (-\infty,0],\7R$. In any case $0\in \mbox{Sp}(P_+)$. 
$P_-$ is treated similarly. For $K$ there is no prefered normalization,
since shifting its origin does not affect the relations (\ref{R1}-\ref{R3}). 
(This is just the fact that the \poinc\ group in 1+1 dimensions has non-trivial one 
dimensional representations $(\Lambda,a)\mapsto e^{iC\Lambda}, C\in\7R$.)
(\ref{R3}) is not affected by shifting $P_\pm$, thus the constant $C$ cannot be
changed. (This reflects the fact that the Lie algebra cohomology $H^2(\6P,\7R)$
is one dimensional.) We have a true representation of the \poinc\ group iff $C$ vanishes.
Now we observe that $C\ne 0$ implies $\mbox{Sp}(P_+)=\mbox{Sp}(P_-)=\7R$, cf.\ eg.\
\cite{bau}. (For $\lambda\in\mbox{Sp}(P_-)$ the differentiated commutation relation
$e^{iaP_+}P_-e^{-iaP_+}=P_-+Ca\11$ implies $\lambda+Ca\in\mbox{Sp}(P_-)\forall a\in\7R$.
Since $\mbox{Sp}(P_-)$ cannot be emply $C\ne 0$ implies $\mbox{Sp}(P_-)=\7R$.)
Before we know that $C=0$ it makes, of course, no sense to consider the joint spectrum 
of $P_+$ and $P_-$. We will however prove that $P_+\ge 0, P_-\ge 0$. By the above this
then entails $C=0$ and positivity in the usual sense: $\mbox{Sp}(P^\mu)\subset\ol{V^+}$.

We consider only $P_+$ since the argument for $P_-$ is the same. The proof is modeled
on the one \cite{dhr4} for DHR sectors. There are simplifications since we are
dealing with soliton {\it auto}morphisms, thus $\bar{\rho}=\rho^{-1}$ is a 
$\alpha_+$-covariant soliton automorphism, too. Yet, we spell the proof out
since there we have to use lightlike instead of spacelike clustering.
The spectra of $P_+^\rho, P_+^{\bar{\rho}}$ containing $0$, positivity of 
$P_+^\rho, P_+^{\bar{\rho}}$ follows from positivity in the vacuum sector if we prove
\be \mbox{Sp}(P_+^\rho)+\mbox{Sp}(P_+^{\bar{\rho}})\subset\mbox{Sp}(P_+^{\pi_0}). 
   \label{add}\ee
Let $\2N_1, \2N_2$ be arbitrary open sets in $\7R$ intersecting 
$\mbox{Sp}(P_+^\rho),\mbox{Sp}(P_+^{\bar{\rho}})$, respectively. Then there is a 
vector $\Psi_1\ne 0$ with $P_+^\rho$-support in $\2N_1$ and, by \cite[Lemma 5.1]{dhr4}, 
a $B\in\2A$ such that $\Psi_2=B\Omega\ne 0$ has $P_+^{\bar{\rho}}$-support in $\2N_2$. 
Now $\Psi_a=\rho(B)e^{-iaP_+^\rho}\Psi_1$ has $P_+^{\pi_0}$-support in $\2N_1+\2N_2$ 
for all $a\in\7R$ and we are done if there is an $a$ such that $\Psi_a\ne 0$. With
\be \|\Psi_a\|^2 = (\Psi_1,e^{iaP_+^\rho}\rho(B^*B)e^{-iaP_+^\rho}\Psi_1)=
   (\Psi_1,\rho(\alpha_{+,a}^{\pi_0}(B^*B))\Psi_1) \ee
such an $a$ exists if the second step in the computation
\be \lim \|\Psi_a\|^2 =\lim(\Psi_1,\rho(\alpha_{+,a}^{\pi_0}(B^*B))\Psi_1)
  \stackrel{?}{=}\|\Psi_1\|^2 \cdot \omega_0(B^*B)=\|\Psi_1\|^2 \cdot \|\Psi_2\|^2 \ne 0 
\label{AA}\ee
is justified for $a\rightarrow +\infty$ or $a\rightarrow -\infty$. The cluster result
\cite[Prop.\ 1.2]{dri0} for lightlike translations gives us weak convergence:
\be \mbox{w}-\lim_{|a|\rightarrow\infty} \alpha_{+,a}^{\pi_0}(B^*B)=\omega_0(B^*B)\,\11
   \quad\forall B\in\2A. \ee
(This result uses that $P_+^{\pi_0}$ has half-sided spectrum and the Reeh-Schlieder 
theorem in the vacuum representation.) Assume that the \sa\ $\rho$ is localized in a 
left wedge. Then it acts normally on the algebras of all right wedges. If $B$ is 
localized in a bounded region then there is a right wedge $W$ such that 
$\alpha_{+,a}(B^*B)\in\2A(W)\ \forall a\ge 0$ and (\ref{AA}) holds for 
$a\rightarrow +\infty$. The maps 
$B\mapsto(\Psi_1,\rho(\alpha_{+,a}^{\pi_0}(B^*B))\Psi_1)$ being
uniformly bounded and the stricly local operators being norm dense in $\2A$, (\ref{AA})
holds for all $B\in\2A$ and $a\rightarrow +\infty$ and we are done.
If $\rho$ is right-localized then let $a\rightarrow -\infty$. \qed

\sectreset{The Solitons of Theories with the Local Fock Property}\label{locfock}
As an application of Thm.\ \ref{main0} we will give in this section a new construction 
of the soliton sectors of the superrenormalizable \qfts\ with the local Fock property, 
like the $P(\phi)_2$ theory (the polynomial $P$ is assumed even in order to have $\7Z_2$
symmetry) or the Yukawa$_2$ model. We briefly indicate the main steps
of the hamiltonian approach to the construction of these models. One begins with a 
finite number of massive free fields and a formal interaction hamiltonian $H_I$. 
We are interested in the case where the non-interacting theory has a group $G$ of inner 
symmetries which leave $H_I$ formally invariant and which survives the renormalization 
(i.e.\ there are no anomalies) but which may be spontaneously broken. Furthermore, we 
assume that $G$ commutes with the \poinc\ action on the free and the interacting theory.
In the rigorous construction of the models, which we sketch on the example of the
$P(\phi)_2$ model, one first obtains the interacting field $\tilde{\phi}$ on the Fock 
space of the free field $\phi_0$ (as a quadratic form) via
\be \tilde{\phi}(x,t)=U^I(t)\,\phi_0(x,0)\,U^I(t)^* \label{propa}\ee
for $x$ in an interval $I$, where $U^I$ is a propagator obtained by smoothly cutting 
off the interaction outside $I$. (In particular $\tilde{\phi}(x,0)=\phi_0(x,0)$.)
One can show that the field $\tilde{\phi}(x,t)$ is independent of the form of the 
cut off of the interaction provided $(x,t)$ is contained in the double cone with basis 
$(I,t=0)$. $\tilde{\phi}$ carries an action of the \poinc\ group, but the action is not
unitarily implemented and there is no invariant vacuum vector. Still there is an 
invariant vacuum state $\omega_0$, and by GNS construction one obtains the physical 
representation $\pi$ of the quasilocal algebra $\2F$ and a unitary representation of
the \poinc\ group. In restriction to arbitrary double cones $\omega_0$ is normal which 
implies local normality of $\pi$. This in turn implies the existence of the physical 
field $\phi(x,t)$. Quantum field theories which can be constructed in the above way 
(including renormalizations of the hamiltonian where necessary) are said to possess the 
{\it local Fock property} \cite{gj}. This property is the only piece of information on 
these models which we will need since (\ref{propa}) in conjunction with the local 
normality of $\pi$ allows to carry over many properties from the free fields to the 
interacting ones. 

\blemma The free massive fields permit local implementers $U^\Lambda_g$
of the unbroken symmetries with the property required in Thm.\ \ref{main0}. 
The $U^\Lambda_g$ can be chosen such that $\alpha_k(U^\Lambda_g)=U^\Lambda_{kgk^{-1}}$
and $U^\Lambda_g U^\Lambda_h=U^\Lambda_{gh}$. \label{L2}\elemma
\prf Unbroken symmetries are unitarily implemented, and the split property for wedges 
implies the existence of disorder operators \cite{mue1,fro2} $U^\2O_L(g)$
implementing the $\alpha_g$ on the left complement of $\2O$, acting trivially on the 
right complement and implementing an even automorphism of $\2F(\2O)$. The latter is
even since the disorder operators commute with $V$.
Now let $\2O\subset\subset\hat{\2O}$ and let $\2O_1, \2O_2$ be as in Fig.\ \ref{rc}. 
Clearly, $U^\Lambda(g)=U^{\2O_1}_L(g) {U^{\2O_2}_L(g)}^*$ acts trivially on $\hat{\2O}'$,
implements $\alpha_g$ on $\2O$ and maps the algebras $\2F(\2O_i), i=1,2$ into
themselves. By duality, $U^\Lambda\in\2F(\hat{\2O}')'=\2F(\hat{\2O})$. Furthermore,
$\alpha_k(U^\Lambda_g)=U^\Lambda_{kgk^{-1}}$ follows from 
$U(k)U^{\2O_i}_L(g)U(k)^*=U^{\2O_i}_L(kgk^{-1}),\ i=1,2$ \cite{mue1} and we have
\bea U^\Lambda_g U^\Lambda_h &=& U^{\2O_1}_L(g) {U^{\2O_2}_L(g)}^*\,U^{\2O_1}_L(h) 
   {U^{\2O_2}_L(h)}^* \nn\\
   &=& U^{\2O_1}_L(g)U^{\2O_1}_L(h) U^{\2O_2}_L(h^{-1}g^{-1}h)U^{\2O_2}_L(h^{-1}) \\
   &=& U^{\2O_1}_L(gh) {U^{\2O_2}_L(gh)}^* = U^\Lambda_{gh}, \nn\eea
where we have used $U^{\2O_2}_g\in\2F(W^{\2O_1}_{LL})''$. \qed\\
\rem The unitaries $U^\Lambda_g$ can be shown to satisfy and to be determined by
$U^\Lambda_g AB\eta=\alpha_g(A)B\eta\ \forall A\in\2F(\2O), B=\2F(\hat{\2O})'$. Here
$\eta$ is the unique vector in $\2P^\natural(\2F(\2O)\vee\2F(\hat{\2O})',\Omega)$
implementing the state
$\omega_\eta(AB)=\omega_0(A)\omega(B), A\in\2F(\2O), B=\2F(\hat{\2O})'$,
where $\omega$ in turn is the product state (existing due to the SPW) which restricts to
$\omega_0$ on $\2F(W^{\hat{\2O}}_{LL})$ and on $\2F(W^{\hat{\2O}}_{RR})$. Recall that 
the usual localized implementer \cite{dl,bdl} is obtained by replacing the product state 
$\omega$ by $\omega_0$.

We summarize the properties of the interacting theory obtained as indicated above.
\blemma Theories with local Fock property satisfy the assumptions 1-5 of Sect.\ 
\ref{General} in their vacuum sector.\label{L1}\elemma 
\prf All these properties are fulfilled by the free scalar and Dirac fields of nonzero 
mass \cite{araki,bu,sum} as well as by theories of finitely many such fields, since they
follow from twisted duality and the split property for wedges \cite{mue3} which is known
to be fulfilled. (Apart from perhaps condition 5 all these properties have been known
before.) The last four properties are of a purely local
nature, thus they carry over immediately to the interacting theory by the local Fock
property. The nontrivial fact that also duality, which is a global property, carries 
over to the interacting theory has been proven in \cite{dri2} for local nets and in 
\cite[Thm.\ 2.8]{sum} for the twisted case. \qed

\blemma The theories with local Fock property permit local implementers of their inner
symmetries with the property required in Thm.\ \ref{main0} and satisfying in addition
$\alpha_k(U^\Lambda_g)=U^\Lambda_{kgk^{-1}},\ U^\Lambda_g U^\Lambda_h=U^\Lambda_{gh}$.
\label{L3}\elemma
\prf By \poinc\ covariance and the fact that $G$ and $\2P$ commute it suffices to prove
the existence of \sa s for double cones $\2O$ which have as basis an interval $I$ in 
the line $t=0$. Let $\2O\subset\subset\hat{\2O}$ with respective bases 
$I\subset\subset\hat{I}$. In view of $\2F_0(\2O)=\2F(I)=\tilde{\2F}(\2O)$ a local 
implementer $U_g^\Lambda$ be the free field (provided by the Lemma \ref{L2}) also 
implements $\alpha_g$ on $\tilde{\2F}(\2O)$. On the Fock space the group $G$ is unitarily
implemented, and since the disorder operators for the free field behave covariantly,
the localized implementers have all desired properties also in the physical
representation $\pi$. \qed

Putting things together we obtain our main result.
\btheor Irreducible \qfts\ in $1+1$ dimensions which have the local Fock property w.r.t.\
a finite number of free massive scalar and/or Dirac fields admit \poinc\ covariant 
locally normal \sa s which can be chosen even and bosonic and satisfying properties 
(A) and (B). \label{main}\etheor
\prf In view of Lemmas \ref{L1} and \ref{L3} the existence of even bosonic \sa s
follows from Thm.\ \ref{main0} and the \poinc\ covariance from Thm.\ \ref{p-covar}.
Properties (A) and (B) are a consequence of Prop.\ \ref{AB}. \qed\\
\rem After we proved the above theorem we discovered that the essential idea of the
existence part is already contained in \cite[pp. 402-4]{fro2}. Still, our results go 
beyond those of \cite{fro2} in several respects. Thm.\ \ref{main0} provides a
convenient sufficient {\it and necessary} condition for the existence of \sa s, which 
also applies to the twisted sectors of holomorphic conformal theories. Furthermore, in 
our proof of \poinc\ covariance we do not appeal to any result of constructive QFT 
except the local Fock property, which, to be sure, is a rather deep result. Finally, 
the fact that one finds \sa s with the covariance (A) and homomorphism (B) properties is
new.

\sectreset{Summary and Outlook}
Thm.\ \ref{dual} completes the abstract treatment of the duality between massive 
theories (satisfying twisted duality and the SPW) with unbroken compact abelian and 
broken abelian symmetry groups, respectively, which was begun in \cite{mue1}. 
The central theme, however, of this work was the observation that \sa s of
massive \qfts\ in $1+1$ dimensions and twisted sectors of holomorphic conformal field 
theories can be treated on equal footing. The crucial property shared by these 
apparently unrelated classes of models is the condition 4 in Sect.\ \ref{General}. 
Sect.\ \ref{locfock} finally complements and extends the earlier rigorous works on 
solitons \cite{fro1,fro2,fre1,fre2,schl2} by establishing sufficiency of the local Fock 
property for the existence of \poinc\ covariant \sa s. The results of that section
make plain that contrary to a widespread belief a profound understanding of the free 
fields is quite useful, if not necessary, in the study of interacting models, 
at least of those with the local Fock property.

Our analysis relies on several deep results from the theory of automorphisms of von 
Neumann algebras. The theorems of Connes/Takesaki and Kallmann/Moore, used in
Thm.\ \ref{dual} and Thm.\ \ref{p-covar}, respectively, make statements on the existence
of continuous unitary group representations inside von Neumann algebras. Crucial 
were furthermore the facts that unital normal endomorphisms of type $I$ factors are 
inner, and that inner tensor-product automorphisms factorize into inner automorphisms.

As emphasized the results of Sect.\ \ref{General} apply also to conformal theories. This
will be the basis for a rigorous analysis of conformal orbifold models in \cite{mue6}.
As mentioned in the Introduction the present analysis was partially motivated by the
desire to understand why the chiral Ising model \cite{ms,boc1} does not fit into the 
analysis of orbifold models given in \cite{dvvv}. As will be discussed further in 
\cite{mue6} this is due to the fact that real fermions on the circle do not admit \sa s 
whereas the existence of the latter -- called `twisted sectors' -- is implicitly assumed
in \cite{dvvv}. (In \cite{kt} the existence of twisted sectors is proved for a class of
WZNW models.) A chiral theory of complex fermions or, more generally, of an even number
of real fermions does possess \sa s which is why for these theories the fusion rules of 
the $\7Z_2$ orbifold theory \cite{boc2} are given by $\7Z_2\times\7Z_2$ or $\7Z_4$ in 
accordance with \cite{dvvv}.

As to solitons in massive models, it would be very interesting to have a proof, 
to the largest possible extent model independent, of bounds on the soliton mass of the
sort proven in \cite{bfg}. Furthermore, one should try to extend Thm.\ \ref{main} to
more general models which do not possess the local Fock property.

\vspace{1cm}
\noindent{\it Acknowledgments.} I would like to thank K.~Fredenhagen, D.~Guido, 
R.~Longo, K.-H.~Rehren and J.~Roberts for useful discussions concerning this work. 
I am particulary indebted to D.~Guido for pointing out an error in the proof of
Thm.\ \ref{p-covar} in an earlier version of this paper.


\begin{thebibliography}{99}
\bibitem{araki} Araki, H.: A lattice of von Neumann algebras associated with the 
  quantum theory of a free Bose field. J. Math. Phys. {\bf 4}, 1343-1362 (1963)
\bibitem{bau} Baumann, K.: Quantum fields in 1+1 dimension carrying a true ray
   representation of the \poinc\ group. Lett. Math. Phys. {\bf 25}, 61-73 (1992)
\bibitem{bfg} Bellissard, J., Fr\"ohlich, J., Gidas, B.: Soliton mass and surface 
   tension in the $(\phi^4)_2$ quantum field model. \cmp {\bf 60}, 37-72 (1978)
\bibitem{boc1} B\"ockenhauer, J.: Localized endomorphisms of the chiral Ising model.
   \cmp {\bf 177}, 265-304 (1996)
\bibitem{boc2} B\"ockenhauer, J.: An algebraic formulation of level one 
   Wess-Zumino-Witten models. \rmp {\bf 8}, 925-947 (1996)
\bibitem{bgl} Brunetti, R., Guido, D., Longo, R.: Modular structure and duality in QFT.
   \cmp {\bf 156}, 201-219 (1993)
\bibitem{bu} Buchholz, D.: Product states for local algebras. \cmp {\bf 36}, 287-304 
   (1974)
\bibitem{bdl} Buchholz, D., Doplicher, S., Longo, R.: On Noether's theorem in \qft.
   Ann. Phys. {\bf 170}, 1-17 (1986)
\bibitem{bdlr} Buchholz, D., Doplicher, S., Longo, R., Roberts, J.E.: A new look at
   Goldstone's theorem. \rmp {\bf Special Issue}, 49-83 (1992)
\bibitem{bmt} Buchholz, D., Mack, G., Todorov, I.: The current algebra on the circle
   as a germ of local field theories. \npb (Proc. Suppl.){\bf 5B}, 20-56 (1988)
\bibitem{fabio} Ciolli, F.: {\it Simmetrie di gauge spontaneamente rotte: Tracce
   osservabili e strutture matematiche associate}. Unpublished diploma thesis, Rome 1997
\bibitem{cole} Coleman, S.: There are no Goldstone bosons in two dimensions.
   \cmp {\bf 2}, 259-264 (1966)
\bibitem{dvvv} Dijkgraaf, R., Vafa, C., Verlinde, E., Verlinde, H.: The operator
   algebra of orbifold models. \cmp {\bf 123}, 485-527 (1989)
\bibitem{dhr2} Doplicher, S., Haag, R., Roberts, J. E.: Fields, observables and
   gauge transformations II. \cmp {\bf 15}, 173-200 (1969)
\bibitem{dhr4} Doplicher, S., Haag, R., Roberts, J. E.: Local observables and particle
   statistics II. \cmp {\bf 35}, 49-85 (1974)
\bibitem{dopl} Doplicher, S.: Local aspects of superselection rules.
   \cmp {\bf 85}, 73-86 (1982)
\bibitem{dl} Doplicher, S., Longo, R.: Standard and split inclusions of von
   Neumann algebras. Invent. Math. {\bf 75}, 493-536 (1984)
\bibitem{dr2} Doplicher, S., Roberts, J. E.: Why there is a field algebra with a 
   compact gauge group describing the superselection structure in particle physics. 
   \cmp {\bf 131}, 51-107 (1990)
\bibitem{dri0} Driessler, W.: Comments on lightlike translations and applications in
   relativistic quantum field theory. \cmp {\bf 44}, 133-141 (1975)
\bibitem{dri1} Driessler, W.: On the type of local algebras in \qft. 
   \cmp {\bf 53}, 295-297 (1977)
\bibitem{dri2} Driessler, W.: Duality and absence of locally generated superselection
   sectors for CCR-type algebras. \cmp {\bf 70}, 213-220 (1979)
\bibitem{fre1} Fredenhagen, K.: Generalizations of the theory of superselection
   sectors. {\it In}: \cite{K}
\bibitem{fre2} Fredenhagen, K.: Superselection sectors in low dimensional \qft.
   J. Geom. Phys. {\bf 11}, 337-348 (1993)
\bibitem{fro1} Fr\"ohlich, J.: New super-selection sectors (`Soliton-states') in 
   two-dimensional Bose quantum field models. \cmp {\bf 47}, 269-310 (1976)
\bibitem{fro2} Fr\"ohlich, J.: Quantum theory of non-linear invariant wave (field)
   equations. Or: Super selection sectors in constructive quantum field theory.
   {\it In:} G. Velo, A. S. Wightman (eds.): {\it Invariant wave equations.} 
   Proceedings Erice 1977
\bibitem{gafr} Gabbiani, F., Fr\"ohlich, J.: Operator algebras and conformal field 
   theory. \cmp {\bf 155}, 569-640 (1993)
\bibitem{glr} Ghez, P., Lima, R., Roberts, J. E.: The spectral category and the Connes
   invariant. J. Oper. Theor. {\bf 14}, 129-146 (1985)
\bibitem{gj} Glimm, J., Jaffe, A.: The $(\Phi^4)_2$ QFT without cutoffs III. The physical
   vacuum. Acta Math. {\bf 125}, 204-267 (1970)
\bibitem{gl} Guido, D., Longo, R.: Relativistic invariance and charge conjugation in
   \qft. \cmp {\bf 148}, 521-551 (1992)
\bibitem{haag} Haag, R.: {\it Local Quantum Physics.} 2nd ed., Berlin: Springer Verlag, 
   1996
\bibitem{haga} Haga, Y.: Crossed products of von Neumann algebras by compact groups.
   T\^{o}hoku Math. J. {\bf 28}, 511-522 (1976)
\bibitem{kt} Kac, V., Todorov, I. T.: Affine orbifolds and rational conformal field
   theory extensions of $W_{1+\infty}$. \cmp {\bf 190}, 57-111 (1997)
\bibitem{K} Kastler, D. (ed.): {\it The algebraic theory of superselection sectors. 
   Introduction and recent results.} World Scientific, 1990
\bibitem{lo1} Longo, R.: Simple injective subfactors. Adv. Math. {\bf 63}, 152-171 (1987)
\bibitem{lm} L\"uscher, M., Mack, G.: Global conformal invariance in QFT. \cmp {\bf 41},
   203-234, (1975)
\bibitem{ms} Mack, G., Schomerus, V.: Conformal field algebra with quantum symmetry 
   from the theory of superselection sectors. \cmp {\bf 134}, 139-196 (1990)
\bibitem{mue1} M\"uger, M.: Quantum double actions on operator algebras and orbifold 
   quantum field theories. \cmp {\bf 191}, 137-181 (1998)
\bibitem{mue3} M\"uger, M.: Superselection structure of massive \qfts\ in $1+1$
   dimensions. hep-th/9705019, to appear in \rmp
\bibitem{mue4} M\"uger, M.: On charged fields with group symmetry and degeneracies of
   Verlinde's matrix S. hep-th/9705018, to appear in Ann. Inst. H. \poinc\ B 
   (Phys. Th\'{e}or.)
\bibitem{mue6} M\"uger, M.: On conformal orbifold models and the twisted quantum double.
   In preparation
\bibitem{n} Niedermaier, M. R.: A derivation of the cyclic form factor equation.
   \cmp {\bf 196}, 411-428 (1998) 
\bibitem{khr2} Rehren, K.-H.: Spin-statistics and CPT for solitons. hep-th/9711085
\bibitem{rob} Roberts, J. E.: Spontaneously broken gauge symmetries and superselection
   rules. {\it In:} Gallavotti, G. (ed.): Proc. International School of Mathematical 
   Physics, Camerino 1974   
\bibitem{rob4} Roberts, J. E.: Some applications of dilatation invariance to structural
   questions in the theory of local observables. \cmp {\bf 37}, 273-286 (1974)
\bibitem{rob1} Roberts, J. E.: Local cohomology and superselection rules. 
   \cmp {\bf 51}, 107-119 (1976)
\bibitem{rob2} Roberts, J. E.: Localization in algebraic field theory.
   \cmp {\bf 85}, 87-98 (1982)
\bibitem{schl2} Schlingemann, D.: On the existence of kink-(soliton-) states in \qft.
   \rmp {\bf 8}, 1187-1203 (1996)
\bibitem{stra} Str\v{a}til\v{a}, S.: {\it Modular theory in operator algebras.}
   Abacus Press, 1981
\bibitem{sum} Summers, S. J.: Normal product states for Fermions and twisted duality
   for CCR- and CAR-type algebras with application to the Yukawa$_2$ quantum field
   model. \cmp {\bf 86}, 111-141 (1982)
\end{thebibliography}
\end{document}